\documentclass[sigconf]{acmart}

\renewcommand\footnotetextcopyrightpermission[1]{} 
\usepackage{multirow}
\usepackage{amsfonts}
\usepackage{graphicx}
\usepackage{subfigure}

\AtBeginDocument{%
  \providecommand\BibTeX{{%
    \normalfont B\kern-0.5em{\scshape i\kern-0.25em b}\kern-0.8em\TeX}}}

\setcopyright{acmcopyright}
\copyrightyear{2018}
\acmYear{2018}
\acmDOI{10.1145/1122445.1122456}

\acmPrice{15.00}
\acmISBN{978-1-4503-XXXX-X/18/06}

\begin{document}

\title{Contrastive Learning for Recommender System}

\author{Zhuang Liu}
\email{liuzhuang@buaa.edu.cn}
\orcid{0000-0001-6149-9667}

\affiliation{%
  \institution{Engineering Research Center of ACAT, Ministry of Education, Beihang University}
  \streetaddress{P.O. Box 1212}
  \city{Beijing}
  \country{China}
}

\author{Yunpu Ma}
\email{cognitive.yunpu@gmail.com}
\affiliation{%
  \institution{Ludwig-Maximilians-Universität München\\
Lehrstuhl für Datenbanksysteme und Data Mining}
  \city{München}
  \country{Germany}}

\author{Yuanxin Ouyang}
\email{oyyx@buaa.edu.cn}
\affiliation{%
  \institution{Engineering Research Center of ACAT, Ministry of Education, Beihang University}
  \city{Beijing}
  \country{China}
}

\author{Zhang Xiong}
\email{xiongz@buaa.edu.cn}
\affiliation{%
 \institution{Engineering Research Center of ACAT, Ministry of Education, Beihang University}
 \streetaddress{Rono-Hills}
 \city{Beijing}
 \country{China}}

\renewcommand{\shortauthors}{Zhuang Liu and Yunpu Ma, et al.}

\begin{abstract}
  Recommender systems, which analyze users' preference patterns to suggest potential targets, are indispensable in today's society.
  Collaborative Filtering (CF) is the most popular recommendation model. Specifically, Graph Neural Network (GNN) has become a new state-of-the-art for CF. In the GNN-based recommender system, message dropout is usually used to alleviate the selection bias in the user-item bipartite graph. However, message dropout might deteriorate the recommender system's performance due to the randomness of dropping out the outgoing messages based on the user-item bipartite graph.
  To solve this problem, we propose a graph contrastive learning module for a general recommender system that learns the embeddings in a self-supervised manner and reduces the randomness of message dropout. Besides, many recommender systems optimize models with pairwise ranking objectives, such as the Bayesian Pairwise Ranking (BPR) based on a negative sampling strategy. 
  However, BPR has the following problems: suboptimal sampling and sample bias. We introduce a new debiased contrastive loss to solve these problems, which provides sufficient negative samples and applies a bias correction probability to alleviate the sample bias. We integrate the proposed framework, including graph contrastive module and debiased contrastive module with several Matrix Factorization(MF) and GNN-based recommendation models. Experimental results on three public benchmarks demonstrate the effectiveness of our framework.
\end{abstract}



\maketitle

\section{Introduction}

To alleviate information overload on the web, recommender system has been widely deployed to perform personalized information filtering \cite{DBLP:conf/sigir/0001DWLZ020}, such as E-commerce platforms (Alibaba, Amazon), social networks (Facebook, Weibo), lifestyle apps (Yelp, Meituan), and so on \cite{Wang2019Neural}.

The recommender system's underlying principle is to estimate how likely a user will adopt an item based on historical interactions like purchases and clicks. CF is the most popular recommendation model, it uses matrix factorization to capture the complex relations between users and items. But the inner product is insufficient to obtain a better performance. Therefore, GNN is widely used for recommender systems since most of the information essentially has a graph structure, and GNN has superiority in representation learning. For example, the interactions between users and items can be considered as the bipartite graph, and the item transitions in sequences can be constructed as graphs as well.
Most prominent among these recent advancements in the GNN \cite{Wang2019Neural,DBLP:conf/aaai/ChenWHZW20,DBLP:conf/sigir/0001DWLZ020}, which can leverage both content information as well as graph structure. GNN-based methods have set a new standard on countless recommender system benchmarks \cite{DBLP:journals/debu/HamiltonYL17,DBLP:conf/kdd/YingHCEHL18}.

\begin{figure}[htbp]
  \centering
  \includegraphics[width=\linewidth]{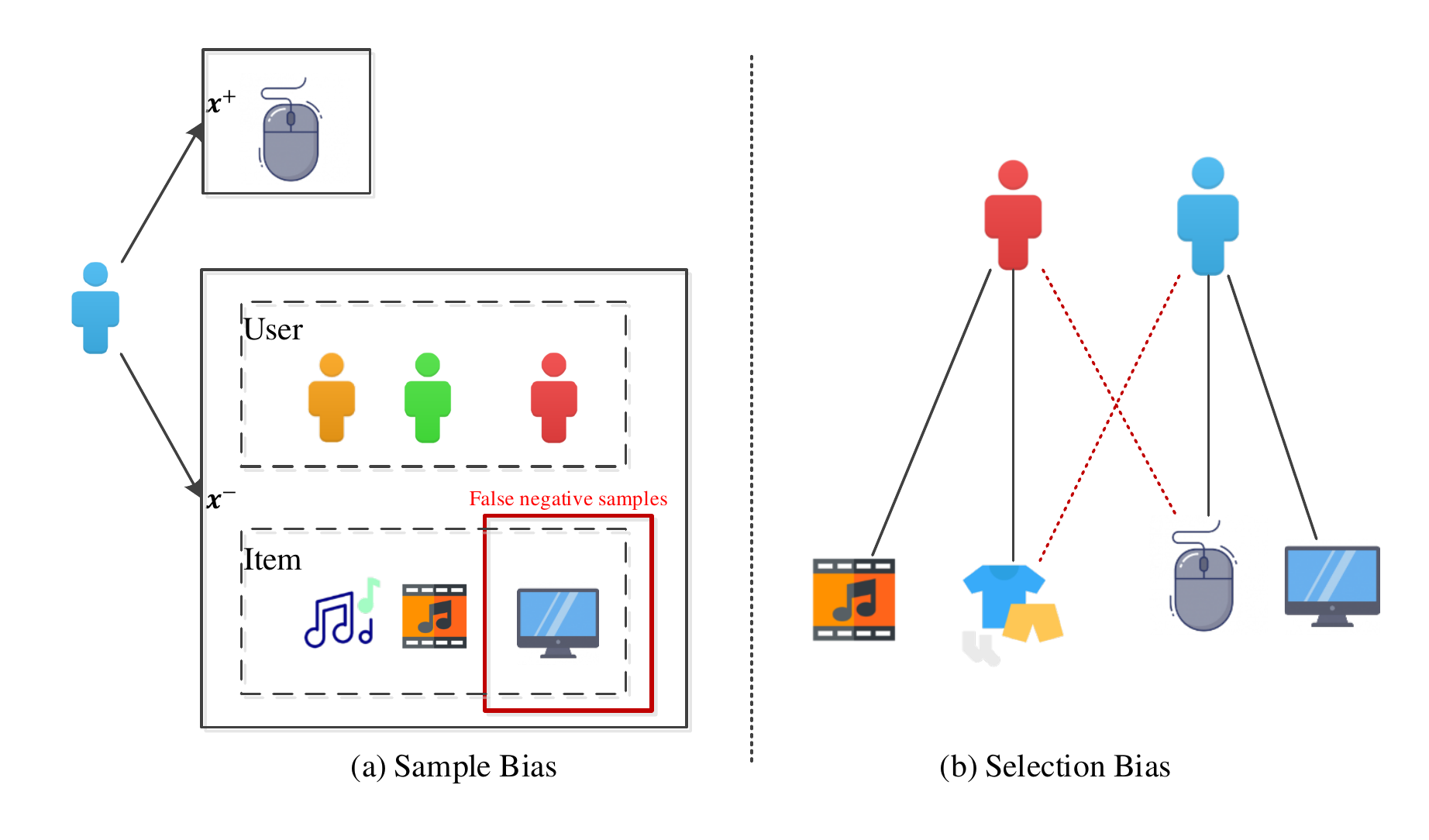}
  \caption{An example of sample bias and selection bias.}
  \label{fig:example}
\end{figure}

For GNN-based collaborative filtering, the user-item bipartite graph is constructed according to the user's interaction with items. GNN is employed as an encoder to learn the representation of users and items.
However, noisy edges are unavoidable in this setting. For example, when a user interacts with a huge amount of items, the irrelevant items may wash out the user's real interest.
As is shown in Figure \ref{fig:example}(b), the (red) user may not like the mouse, but due to the position bias, the mouse ranked highly causes the user to collect; this will be a \emph{selection bias}. To alleviate the selection bias, the commonly used method is message dropout \cite{Wang2019Neural,DBLP:conf/sigir/0001DWLZ020}, which randomly drops out the outgoing messages based on the user-item graph. However, this may drop some essential information, resulting in a decrease in recommendation performance (see subsection \ref{sec:ablation}, the message dropout cause the worse performance in Steam dataset.)

Therefore, to reduce selection bias and obtain a better representation for each user and item, we propose a graph contrastive learning module, which learns users' and items' embedding in a self-supervised manner. Specifically, for each node in the user-item graph, random perturbations are applied to its subgraph, namely random dropping a subset of its subgraph edges. Here, we randomly perturb the subgraph of each node twice. And then learn the embedding of users and items by maximizing the similarity between the representations of two randomly perturbed versions of the link structure of the same node's local subgraph. We still use GNN models to produce two representations of the same user (item) and leverage a contrastive learning loss to maximize their similarity.

Besides, many recommender systems optimize models with BPR \cite{DBLP:conf/uai/RendleFGS09} based on a negative sampling strategy. However, the BPR method has the following two problems:

(1) Suboptimal sampling: The BPR method randomly samples unobserved items for each positive item as a negative sample, which is unfair to ignored items and may lead to inferior representations for these items. 


(2) Sample bias: BPR assumes that the observed interactions, which are more reflective of a user's preferences, should be assigned higher prediction values than unobserved ones. Moreover, the negative items are typically taken to be randomly sampled from the unobserved items, implicitly accepting that these items may, in reality, actually have the same label (positive items). This phenomenon, which we refer to as sample bias, can empirically lead to a significant performance drop \cite{DBLP:journals/corr/abs-2007-00224}. As is shown in Figure \ref{fig:example}(a), a user is keen on electronic products. When randomly sampling, items that are consistent with the user's interest may be collected (the computer in the red box in Figure \ref{fig:example}(a)); this will be a false negative sample.



To tackle the above problems, in this paper, we introduce a new loss called the debiased contrastive loss to obtain the refined embedding for each user and item. In particular, given a new batch to process, we take all the users and its positive item of other instances in the current batch as the negative examples. Here, we also treat the users as negative samples since one of the key properties - uniformity in contrastive loss prefers a feature distribution that preserves maximal information, namely if features of a class are sufficiently well clustered, they are linearly separable with the rest feature space \cite{DBLP:journals/corr/abs-2005-10242}. In this way, we provide sufficient negative samples to achieve a satisfying performance.
Moreover, it can efficiently incorporate these negative samples per batch, thanks to computation reuse. However, these negative samples still contain a few positive items, as shown in the red box in Figure \ref{fig:example}(a). To solve this problem, we define bias correction probability $\tau^+$ to alleviate these sample bias; namely, there is a $\tau^+$ probability that they are positive items among these negative samples.

To summarize, this work makes the following main contributions:
\begin{itemize}
\item We propose a graph contrastive learning module for general GNN-based recommender systems to learn the representation of users and items in a self-supervised manner.
\item We develop a general debiased contrastive loss for recommender systems that improve suboptimal sampling and corrects sampling bias.  
\item We treat the users as negative samples to prefer a feature distribution that preserves maximal information. To the best of our knowledge, this is the first study to consider this sample strategy.
\item We conduct empirical studies on three public datasets. Extensive results demonstrate the state-of-the-art performance of our method and its effectiveness in each component.
\end{itemize}

\section{Related Work}
In this section, we will briefly review several lines of works closely related to ours, including graph-based recommendation and contrastive learning.

\textbf{Graph-based Recommendation}.
Graph-based recommendation methods mainly exploit the user-item interaction graph to infer user preference. Early works \cite{DBLP:conf/ijcai/GoriP07,DBLP:journals/tkde/HeGKW17} used label propagation to capture the collaborative filtering (CF) effect. They define the labels as a user's interacted items and propagate the labels on the user-item graph. Besides, HOP-Rec \cite{DBLP:conf/recsys/YangCWT18} applies random walks on the user-item interaction graph to derive similarity scores for user-item pairs. The superior performance of HOP-Rec is that incorporating the connectivity information is beneficial to obtain better embeddings in capturing the CF effect. Recently, some other methods have been devised by devising a specialized graph convolution operation on the user-item interaction graph. For example, GC-MC \cite{DBLP:journals/corr/BergKW17} applies the Graph Convolution Network (GCN) on the user-item graph; it employs one convolutional layer to exploit the direct connections between users and items. PinSage \cite{DBLP:conf/kdd/YingHCEHL18} is an industrial solution that employs multiple graph convolution layers on the item-item graph for Pinterest image recommendation. NGCF \cite{Wang2019Neural} exploits the user-item graph structure by propagating embeddings on it; this leads to the expressive modeling of high-order connectivity in the user-item graph. LightGCN \cite{DBLP:conf/sigir/0001DWLZ020} is the light version of NGCF, which including only the most essential component in GCN - neighborhood aggregation - for collaborative filtering. It is a state-of-the-art GCN-based recommender model.

\textbf{Contrastive Learning}.
Contrastive learning, which aims to learn high-quality representation via a self-supervised manner, recently achieves remarkable successes in various fields\cite{DBLP:conf/cvpr/He0WXG20,DBLP:conf/iclr/HjelmFLGBTB19,DBLP:journals/corr/abs-2002-05709}. The common motivation behind these work is the InfoMax principle \cite{DBLP:journals/computer/Linsker88}, which we here instantiate as maximizing the mutual information (MI) between two views \cite{DBLP:conf/nips/BachmanHB19}. It learns discriminative representations by contrasting positive and negative samples. In natural language processing, the most classic model - Word2vec \cite{DBLP:conf/nips/MikolovSCCD13} uses co-occurring words and negative sampling to learn word embeddings. To efficiently learn sentence representations, \cite{DBLP:conf/iclr/LogeswaranL18} treat the context sentences as positive samples and the others as negative samples to optimize a conrastive loss. In computer vision, a large collection of work \cite{DBLP:conf/cvpr/HadsellCL06,DBLP:conf/cvpr/He0WXG20,DBLP:journals/corr/abs-1906-05849,DBLP:conf/cvpr/WuXYL18} learns self-supervised image representation by minimizing the distance between two views of the same image. 

Although contrastive learning has achieved relatively remarkable results in natural language processing and computer vision, there is still little work of contrastive learning in recommender systems. Therefore, in this paper, we propose a general contrastive learning framework for recommender systems. 

\section{Preliminary}
\subsection{Problem Setup}
Let $U=\{u_1,u_2,\cdots,u_n\}$ and $I=\{i_1,i_2,\cdots,i_m\}$ be the sets of users and items respectively, where $n$ is the number of users, and $m$ is the number of items. We assume that $G\in\{0,1\}^{n\times m}$ is the undirected bipartite user-item graph, which contains two disjoint node sets, users $U$ and items $I$. If $u$ gives a rating to $i$, $g_{ui}=1$, otherwise we employ 0 to represent the unknown rating. An example of user-item graph is shown in Figure \ref{fig:model}(a). Following \cite{Wang2019Neural}, we use an embedding vector $\textbf{e}_u$ to denote a user $u$ and an embedding vector $\textbf{e}_i$ to represent an item $i$, where $d$ is the dimension of embedding vector. Then, the recommendation task asks that, given the user-item graph $G$, our goal is to predict each user's preferences to unobserved items. 

\subsection{GNN for Recommendation}
\label{sec:3.2}
In this paper, we mainly focus on graph neural network recommendation methods. Specifically, our study focuses solely on the pure user-item bipartite graph, which does not exploit auxiliary information such as user profile or item profile, and so on. 

Given the user-item bipartite graph, the critical challenge is propagating the information on the graph structure. There are three main issues to deal with: (1) \textbf{Neighbor Aggregation} How to aggregate the information from neighbor nodes. (2) \textbf{Information Update} How to integrate the central node representation and its neighbors' aggregated representation. (3) \textbf{Final Node Representation} How to combine the node representations in all layers. Here, we introduce several widely used graph neural networks for recommender systems.
Let $\textbf{e}_u^{(k)}$ and $\textbf{e}_i^{(k)}$ respectively denote the refined embedding of user $u$ and item $i$ after $k$ layers propagation, $\mathcal{N}_u$ denotes the set of items that are interacted by user $u$, $\mathcal{N}_i$ denotes the set of users that interact with item $i$, $\textbf{W}_1$, $\textbf{W}_2$, $\textbf{W}_3$ and $\textbf{W}_4$ are trainable weight matrix, $K$ is the layers of GNN and $\sigma$ is the nonlinear activation function. Due to the space limitation, we take the user node as examples.

    (1) \textbf{GC-MC}: It first applies GCN to encoder the user-item bipartite graph, and then uses a bilinear decoder to reconstruct links in the bipartite interaction graph: 
    \begin{equation}
    \label{equation:gcmc}
        \textbf{e}_u^{(k+1)}=\textbf{W}_1\sigma\left(\sum_{i\in\mathcal{N}_u}\frac{1}{\sqrt{|\mathcal{N}_u|}\sqrt{|\mathcal{N}_i|}}\left(\textbf{W}_2\textbf{e}_i^{(k)}\right)\right),
    \end{equation}
    Here, GC-MC only consider the first-order neighbors, that is $K=1$.
    
    (2) \textbf{PinSage}: Inspired by the GraphSage method, PinSage concatenates the aggregated neighbor vector with $u$'s current representation. The normalization operation controls the scale of the node embedding and makes training more stable:
    \begin{equation}
    \label{equation:pinsage}
        \textbf{e}_u^{(k+1)}=\sigma\left(\textbf{W}_1\left(\textbf{e}_u\left|\right|\sum_{i\in\mathcal{N}_u}\frac{1}{\sqrt{|\mathcal{N}_u|}\sqrt{|\mathcal{N}_i|}}\left(\textbf{W}_2\textbf{e}_i^{(k)}\right)\right)\right),
    \end{equation}
    \begin{equation}
        \textbf{e}_u^{(k+1)}=\textbf{e}_u^{(k+1)}/\left|\left|\textbf{e}_u^{(k+1)}\right|\right|_2
    \end{equation}
    The final node embeddings are the non-linear transformation of the representations in the last layer, i.e., $\textbf{e}_u^*=\textbf{W}_3\cdot\sigma\left(\textbf{W}_4\textbf{e}_u^{(K)}\right)$
    
    (3) \textbf{NGCF}: It leverages the user-item interaction graph to propagate embeddings as :
    \begin{equation}
    \label{equation:ngcf}
        \textbf{e}_u^{(k+1)}=\sigma\left(\textbf{W}_1\textbf{e}_u^{(k)}+\sum_{i\in\mathcal{N}_u}\frac{\textbf{W}_1\textbf{e}_i^{(k)}+\textbf{W}_2\left(\textbf{e}_i^{(k)}\odot\textbf{e}_u^{(k)}\right)}{\sqrt{|\mathcal{N}_u|}\sqrt{|\mathcal{N}_i|}}\right),
    \end{equation}
    After $K$ layers' propagation, NGCF adopts the concatenation strategy to take full advantage of representations in different layers.
    
    (4) \textbf{LR-GCCF}: It removes non-linearity transformation in GCN:
    \begin{equation}
    \label{equation:lr-gccf}
        \textbf{e}_u^{(k+1)}=\textbf{W}_1\left(\textbf{e}_u^{(k)}+\sum_{i\in\mathcal{N}_u}\frac{1}{\sqrt{|\mathcal{N}_u|}\sqrt{|\mathcal{N}_i|}}\textbf{e}_i^{(k)}\right),
    \end{equation}
    
    (5) \textbf{LightGCN}: It is the light version of NGCF, which including only the most essential component in GCN - neighborhood aggregation:
    \begin{equation}
    \label{equation:lightgcn}
        \textbf{e}_u^{(k+1)}=\sum_{i\in\mathcal{N}_u}\frac{1}{\sqrt{|\mathcal{N}_u|}\sqrt{|\mathcal{N}_i|}}\textbf{e}_i^{(k)},
    \end{equation}
    LightGCN changes the way of obtaining final embedding from concatenation (i.e., $\textbf{e}_u^*=\textbf{e}_u^{(0)}\left|\right|\cdots\left|\right|\textbf{e}_u^{(K)}$) to sum (i.e., $\textbf{e}_u^*=\textbf{e}_u^{(0)}+\cdots+\textbf{e}_u^{(K)}$).

\section{Methodology}
In this section, we will first give an overview of the proposed framework, then detail each model component and finally discuss how to learn the model parameters.
\begin{figure}[htbp]
  \centering
  
  \includegraphics[width=\linewidth]{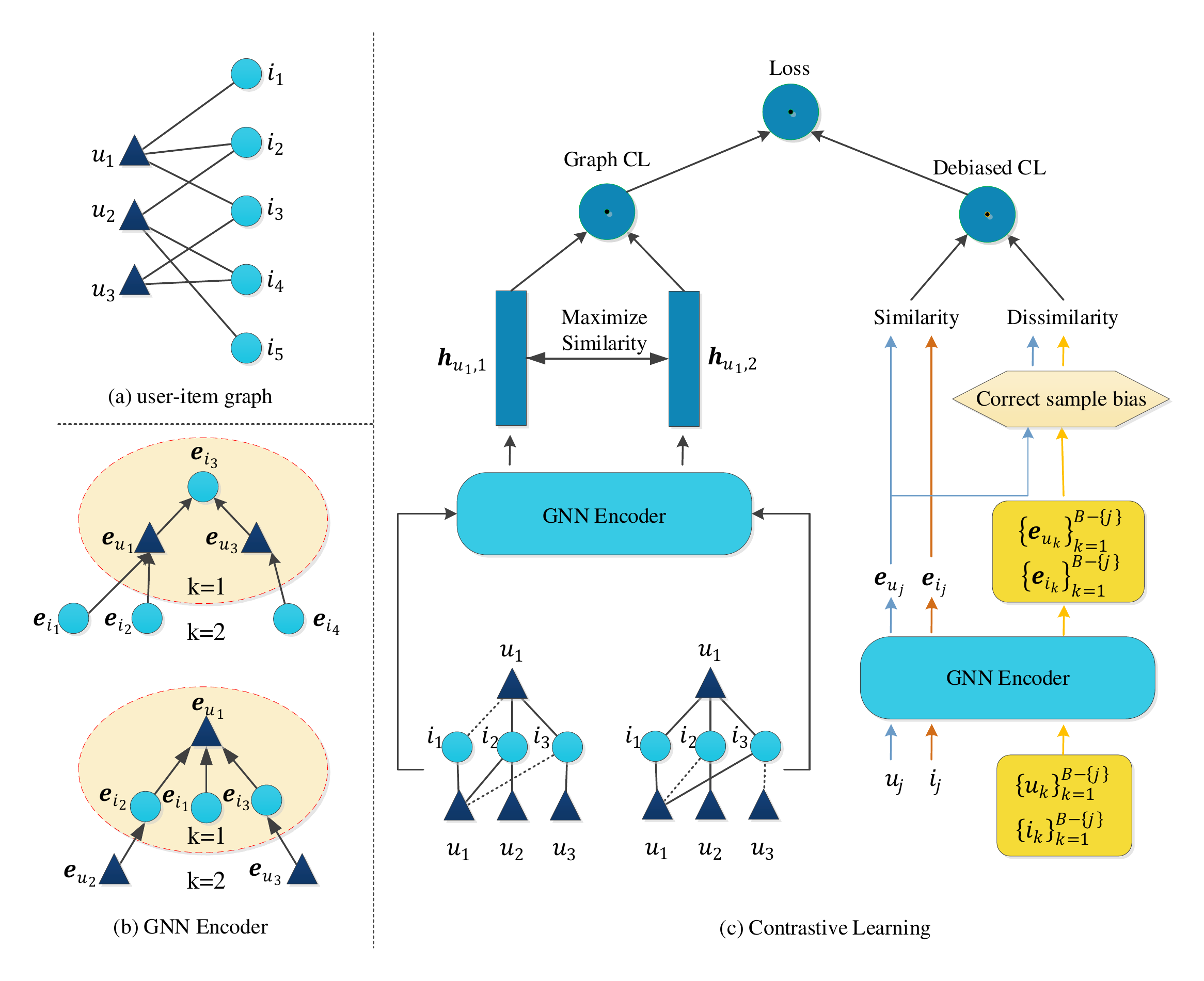}
  \caption{The overall architecture of the proposed framework.}
  \label{fig:model}
\end{figure}

\subsection{An Overview of the Proposed Framework}
The architecture of the proposed framework is shown in Figure \ref{fig:model}. First, we construct an undirected bipartite user-item graph based on the historical interaction of the users. Then, we use graph neural networks such as the GNN models mentioned in subsection \ref{sec:3.2} to encode the user behaviors and items, as is shown in Figure \ref{fig:model}(b). Finally, to better optimize such GNN-based recommendation algorithms, we propose a contrastive learning framework containing a graph contrastive learning module and a debiased contrastive learning module to learn the embeddings in a self-supervised manner as shown in Figure \ref{fig:model}(c). Specifically, graph contrastive learning is to reduce the selection bias and randomness caused by message dropout. It can obtain a better representation for each user and item. Debiased contrastive learning aims to correct the sampling bias and provides sufficient negative samples to achieve a satisfying performance.

\subsection{Graph Perturbation}
Graph perturbation aims to reduce the selection bias and obtain a better representation for each user and item. Its objective is to learn user or item representations by maximizing the similarity between the embeddings of two randomly perturbed versions of the same user's or item's neighborhood subgraph using a graph contrastive loss in the embedding space. 

As shown in the left of Figure \ref{fig:model}(c). First, for a given user-item bipartite graph, we apply a stochastic perturbation to the $L$-hop subgraph centered at each user (item), which results in two neighborhood subgraphs that allow us to obtain two representation of the same user (item), which we consider as positive examples. In this work, the subgraph structure is transformed by randomly dropping edges with probability $p$ using samples from a Bernoulli distribution. Second, we apply several GNN methods mentioned in \ref{sec:3.2} to learn representations of the two transformed $L$-hop subgraphs associated with each user $u$ (item $i$). It is worth noting that the encoder here can be any GNN models. Finally, a graph contrastive loss function is defined to enforce maximizing the consistency between positive pairs $\textbf{h}_{u,1}$, $\textbf{h}_{u,2}$ compared with negative pairs, where $\textbf{h}_{u,1}$, $\textbf{h}_{u,2}$ stand for the representation of two transformations of the $L$-hop subgraph around the same user.

More formally, we randomly sample a minibatch $\mathcal{B}$ containing $B$ users and define their corresponding $L$-hop subgraphs. We apply two transformations to each of the user's subgraphs resulting in $2B$ subgraphs enabling us to get positive pairs of representations for the contrastive learning task. Negative pairs are not explicitly samples but generated from the other $(2B-2)$ examples within the same minibatch as in \cite{DBLP:conf/kdd/ChenSSH17,DBLP:journals/corr/abs-2002-05709}.

For each user $u$ in the minibatch, we compute the following loss function, which is based on a normalized temperature-scales cross entropy \cite{DBLP:conf/nips/Sohn16,DBLP:journals/corr/abs-1807-03748,DBLP:conf/cvpr/WuXYL18}:
\begin{equation}
\label{equation:E10}
    l(u)=l_{1,2}(u)+l_{2,1}(u)
\end{equation}
where $l_{i,j}(u)$ is defined as 
\begin{equation}
\begin{aligned}
\label{equation:E11}
    l_{i,j}(u)=-\log
    \frac{e^{s\left(\textbf{h}_{u,i}\textbf{h}_{u,j}\right)/t_1}}{\sum_{v\in\mathcal{B}}\mathbf{1}_{[v\neq u]}e^{s\left(\textbf{h}_{u,i}\textbf{h}_{v,i}\right)/t_1}+\sum_{v\in\mathcal{B}}e^{s\left(\textbf{h}_{u,i}\textbf{h}_{v,j}\right)/t_1}}
\end{aligned}
\end{equation}
where $s\left(\textbf{h}_{u,i}\textbf{h}_{v,j}\right)=\textbf{h}_{u,i}^\top\textbf{h}_{u,j}/\left|\left|\textbf{h}_{u,i}\right|\right|\left|\left|\textbf{h}_{u,j}\right|\right|$ is the cosine similarity between the two representations $\textbf{h}_{u,i}$ and $\textbf{h}_{u,j}$, $\mathbf{1}_{[u\neq v]}$ is an indicator function equals to 1 iff $u\neq v$ and $t_1$ is a temperature parameter. (It is worth noting that in this work, we only consider the graph contrastive loss for user perturbation. We leave the item perturbation in the future.) Finally, the graph contrastive loss is defined as:
\begin{equation}
    L_{GCL}=\frac{1}{B}\sum_{u\in\mathcal{B}}l(u)
\end{equation}

\subsection{Debiased Contrastive Loss}
As mentioned earlier, BPR loss suffers from problems such as suboptimal sampling and sampling bias in the recommender system. To solve these problems, we propose a self-supervised debiased contrastive learning method, as is shown in the right of Figure \ref{fig:model}(c). 

Given an arbitrary user $u$ (an item $i$), we can obtain its embedding $\textbf{e}_u$ (or $\textbf{e}_i$) based on GNN models. The task of recommender system is to retrieve the top $k$ items relevant to the user $u$ by finding the top $k$ candidate $\{\textbf{e}_i\}_{i\in I}$ similar to $\textbf{e}_u$. Most implementations use inner product or cosine similarity $\phi(u,i)=<\textbf{e}_u,\textbf{e}_i>$ as the similarity score. BPR method is widely used in the learning procedure fits the data. It randomly samples an unobserved item for each positive item as a negative sample, leading to suboptimal sampling mentioned earlier. Therefore, we consider more negative samples for each positive item, name the learning procedure following the contrastive loss \cite{DBLP:journals/corr/abs-1807-03748,DBLP:conf/nips/Sohn16,zhou2020contrastive}:
\begin{equation}
\label{equation:E4}
\begin{aligned}
L_{CL}=\frac{1}{|\mathcal{D}|}\sum_{(u,i)\in\mathcal{D}}-\log\frac{e^{\phi(u,i)}}{e^{\phi(u,i)}+\sum_{l=1}^Le^{\phi(u,i_l)}}.
\end{aligned}
\end{equation}
where $\mathcal{D}=\{(u,i):u\in U, i\in I\quad \text{and}\quad g_{ui}=1\}$ is the training set, $i_l$ is the negative items that the user $u$ have non-interacted, $L$ denotes the number of negative items for $u$.

Generally speaking,  the negative items are randomly sampled from unobserved items of a user for the BPR method. However, this may exist false negative items; namely, some negative items may be potential positive items that the user likes. In order to solve the sample bias, we develop a debiased contrastive loss for recommendation algorithms. We assume the sampled negative items $\{i_l\}_{l=1}^L$ come from the positive items with probability $\tau^+$. That is, in these negative items, we believe that the proportion of true negative items is $1-\tau^+$. Therefore, for each item $i_l$ in $\{i_l\}_{l=1}^L$, we can obtain the negative score for user $u$ as following:

\begin{equation}
\label{equation:E5}
    g\left(u,i,\{i_l\}_{l=1}^L\right)=\frac{1}{1-\tau^+}\left(\frac{1}{L}\sum_{l=1}^Le^{\phi(u,i_l)}-\tau^+e^{\phi(u,i)}\right),
\end{equation}
where the second term in brackets can be regarded as the positive score for user $u$, by dividing by the probability $1-\tau^+$ of the true negative samples, we further scale the negative score. Therefore, we can obtain the debiased contrastive loss as follows:
\begin{equation}
\label{equation:E6}
\begin{aligned}
L_{DCL}=\frac{1}{|\mathcal{D}|}\sum_{(u,i)\in\mathcal{D}}-\log\frac{e^{\phi(u,i)}}{e^{\phi(u,i)}+Lg\left(u,i,\{i_l\}_{l=1}^L\right)}.
\end{aligned}
\end{equation}

Following the sampling strategy in \cite{DBLP:conf/kdd/ChenSSH17}, we consider the other (2B-2) examples in the current minibatch as negative samples. Here, we treat users and items in the current minibatch as negative samples for the following reasons: (1) In GNN models, users and items are the same, and both of them are obtained the corresponding embeddings by fusing their high-order neighbors. Furthermore, it is impossible to recommend a user to the user, so it is reasonable to treat them as negative samples. (2)In \cite{DBLP:journals/corr/abs-2005-10242}, it has shown that the contrastive loss contains two key properties: alignment(closeness) and uniformity. One of them - uniformity prefers a feature distribution that preserves maximal information, i.e., the uniform distribution on the unit hypersphere. Moreover, if a class's features are sufficiently well clustered, they are linearly separable from the rest of the feature space. In recommender systems,  the users and items are different classes; they should separable from each other in the embedding space. This further indicates that we treat the users as negative samples are reasonable.

Therefore, we can modify equation (\ref{equation:E5}) as:
\begin{equation}
\begin{aligned}
\label{equation:E7}
    g\left(u_j,i_j,\{u_k\}_{k=1}^{B-\{j\}},\{i_k\}_{k=1}^{B-\{j\}}\right)=
    \frac{1}{1-\tau^+}\\
    \left(\frac{1}{2B-2}\sum_{k=1}^{B-1}\left(e^{\phi(u_j,u_k)}+e^{\phi(u_j,i_k)}\right)
    -\tau^+e^{\phi(u_j,i_j)}\right),
\end{aligned}
\end{equation}
where $j$ is the index of user or item in a minibatch. $\{u_k\}_{k=1}^{B-\{j\}}$ and $\{u_k\}_{k=1}^{B-\{j\}}$ denotes the other $B-1$ examples in a minibatch for a user and an item.

In this work, we use the cosine similarity with temperature $t_2>0$ as the similarity function, i.e. $\phi(u,i)=\frac{<\textbf{e}_u,\textbf{e}_i>}{t_2}$ assuming that $\textbf{e}_u$ and $\textbf{e}_i$ are both $l2$-normalized. In addition, \cite{DBLP:journals/corr/abs-2007-00224} has proved that the equation (\ref{equation:E7}) exits its theoretical minimum $e^{1/t_2}$. So we constrain the equation (\ref{equation:E7}) to be greater than $e^{1/t_2}$ to prevent calculating the logarithm of a negative number, that is:
\begin{equation}
\begin{aligned}
\label{equation:E8}
    g'\left(u_j,i_j,\{u_k\}_{k=1}^{B-\{j\}},\{i_k\}_{k=1}^{B-\{j\}}\right)=\\
    \max\{g\left(u_j,i_j,\{u_k\}_{k=1}^{B-\{j\}},\{i_k\}_{k=1}^{B-\{j\}}\right),e^{1/t_2}\}.
\end{aligned}
\end{equation}

Finally, the debiased contrastive loss for recommendation algorithm in this work is defined as:
\begin{equation}
\label{equation:E9}
\begin{aligned}
L_{DCL}=\frac{1}{|\mathcal{D}|}\sum_{(u_j,i_j)\in\mathcal{D}}-\log\frac{e^{\phi(u_j,i_j)}}{e^{\phi(u_j,i_j)}+(2B-2)g'}.
\end{aligned}
\end{equation}

\subsection{Model Training}
To effectively learn the parameters of our framework, we need to specify an objective function to optimize. Based on the above two modules: $L_{GCL}$ to reduce the selection bias and obtain a better representation for each user and item. $L_{DCL}$ to correct the sampling bias and provide sufficient negative samples to speed up the training process, we integrate these two parts in an end-to-end fashion; namely, the overall training loss can be rewritten as:
\begin{equation}
    \mathcal{L}=\sum_{(u,i)\in\mathcal{D}}\beta L_{GCL}+(1-\beta) L_{DCL}+\lambda\left(||\textbf{e}_u||_2^2+||\textbf{e}_i||_2^2\right)
\end{equation}
where $\beta$ is a balance coefficient. $\lambda$ controls the $L_2$ regularization strength. We employ mini-batch Adam as the optimizer \cite{DBLP:journals/corr/KingmaB14}. Its main advantage is that the learning rate can be self-adaptive during the training phase, which eases the pain of choosing a proper learning rate.

\section{Experiments}
In this section, we conduct experiments to evaluate the performance of our framework on three public datasets. Specifically, we aim to answer the following four research questions:
\begin{itemize}
\item \textbf{RQ1}: Is our proposed framework applicable to general and benchmark algorithms for recommender systems, such as MF, DMF, NGCF, LightGCN, etc.?

\item \textbf{RQ2}: What is the contribution of various components in our framework, e.g., graph perturbation, debiased contrastive loss, etc.?
\item \textbf{RQ3}: Is our framework more efficient compared with the state-of-the-art model?
\item \textbf{RQ4}: How robustness our framework respect to hyper - parameters, such as bias correction probability $\tau^+$ and graph perturbation rate $p$)?
\end{itemize} 

\subsection{Experimental Settings}
\subsubsection{Datasets} To evaluation the effectiveness of our framework, we conduct experiments on three real-world datasets. These datasets vary significantly in domains, size, and sparsity. 

\begin{itemize}
\item \textbf{Yelp2018}: This dataset is adopted from the 2018 edition of the Yelp challenge. Wherein, the local businesses like restaurants and bars are viewed as the items.
\item \textbf{Amazon-Book}: This is a series of product review datasets crawled from Amazon.com, which are widely used for product recommendation\cite{DBLP:journals/corr/HeM16}. They are split into separate datasets according to the top-level product categories on Amazon. In this work, we adopt a subset, Amazon-Book. 
\item \textbf{Steam}: This is a dataset collected from Steam, a large online video game distribution platform, by Kang and McAuley\cite{2018Self}. The dataset also includes rich information that might be useful in future work, like users' play hours, pricing information, media score, category, etc.
\end{itemize}

For dataset preprocessing, the Yelp2018 and Amazon-Book are the same as \cite{Wang2019Neural}; namely, we retain users and items with at least ten interactions. In the Steam dataset, we following the common practice in \cite{2019BERT4Rec}, keeping users with at least five feedbacks. The statics of the processed datasets are summarized in Table \ref{tab:datasets}. For each dataset, same as \cite{Wang2019Neural}, we randomly select 80\% of historical interactions of each user to constitute the training set and treat the remaining as the test set.
\begin{table}
  \caption{Statistics of the datasets.}
  \label{tab:datasets}
  \begin{tabular}{c|c|c|c|c}
    \toprule
    Dataset & \#Users & \#Items & \#Interactions & Density\\
    \midrule
    Yelp2018 & 31668 & 38048 & 1561406 & 0.00130\\
    Amazon-Book & 52643 & 91599 & 2984108 & 0.00062\\
    Steam & 281428 & 13044 & 3488885 & 0.00095\\
  \bottomrule
\end{tabular}
\end{table}

\subsubsection{Evaluation Metrics} In order to evaluate the quality of the recommendation algorithms, we adopt two widely-used top-$K$ metrics: recall@$K$ and ndcg@$K$, which are computed by the all-ranking protocol - all items that are not interacted by a user are the candidates. Specifically, recall@$K$ measures the number of items that the user likes in the test data that has been successfully predicted in the top-$K$ ranking list. Moreover, ndcg@$K$ considers the hit positions of the items and gives a higher score if the hit items in the top positions. For both metrics, the larger the values, the better the performance. By default, we set $K=20$. We report the average metrics for all users in the test set.

\subsubsection{Baselines} To verify the effectiveness of our framework, we compare it with the following representative baselines:
\begin{itemize}
\item \textbf{MF}\cite{DBLP:journals/corr/abs-1205-2618}: This is matrix factorization optimized by the BPR loss, which exploits the user-item direct interactions only as the target value of interaction function.
\item \textbf{DMF}\cite{ijcai2017-447}: This is a novel matrix factorization model with neural network architecture. It takes the user-item matrix as input, and uses a deep structure learning architecture to learn a common low dimensional space for the representations of users and items.
\item \textbf{GCN} \cite{DBLP:conf/iclr/KipfW17}: It is an typical variant of convolutional neural networks which operate directly on graph-structured data. In this work, we apply it on user-item interaction graph.
\item \textbf{GC-MC} \cite{DBLP:journals/corr/BergKW17}: It is a graph auto-encoder framework based on differentiable message passing on the bipartite user-item interaction graph, where only the first-order neighbors are considered.
\item \textbf{PinSage} \cite{DBLP:conf/kdd/YingHCEHL18}: It is designed to employ GraphSAGE \cite{DBLP:conf/nips/HamiltonYL17} on item-item graph. In this work, we apply it on user-item graph. Especially, we employ two graph convolution layers as suggested in \cite{DBLP:conf/kdd/YingHCEHL18}, and the hidden dimension is set equal to the embedding size.
\item \textbf{NGCF} \cite{Wang2019Neural}: This method integrates the user-item interactions - more specifically, the bipartite graph structure - into the embedding process, which exploits the graph structure by propagating embeddings on it to model the high-order connectivity.
\item \textbf{LR-GCCF} \cite{DBLP:conf/aaai/ChenWHZW20}: It is a residual network structure that is specifically designed for CF with user-item interaction modeling, which alleviates the over smoothing problem in graph convolution aggregation operation with sparse user-item interaction data.
\item \textbf{LightGCN} \cite{DBLP:conf/sigir/0001DWLZ020}: It is a state-of-the-art GCN-based model, which is also the light version of NGCF. It includes only the most essential component in GCN - neighborhood aggregation - for collaborative
\item \textbf{LightGCN-single} \cite{DBLP:conf/sigir/0001DWLZ020}: It is a variant of LightGCN, which does not use layer combination. Namely, it only uses the $K$-th layer output for final prediction for a $K$-layer LightGCN.
\end{itemize}

For fair comparison, all baseline methods optimize the BPR loss as shown in Equation (\ref{equ:bpr}). For each baseline, we retain the base model to encode the user behaviors and items, and change the loss function to GCL+DCL. Here, we use LightGCN to encode the two perturbed sub-graphs for MF and DMF. We define our methods as XX++ (e.g., NGCF++).

\subsubsection{Parameter Settings} We implemented our proposed method on the basis of Pytorch\footnote{\url{https://pytorch.org/}}, a well-known Python library for neural networks. We optimize all models using the Adam optimizer with the Xavier initialization procedure \cite{DBLP:journals/jmlr/GlorotB10}. The embedding size is fixed to 128 and the batch size to 2048 for all baseline models and our algorithm. Grid search is applied to choose the learning rate and the $L_2$ regularization coefficient $\lambda$ over the ranges $\{0.0001,0.001,0.01\}$ and $\{1e^{-6},1e^{-5},\cdots,1e^{-2}\}$, respectively. And in most cases the optimal value is $0.001$ and $1e^{-4}$. We set the layer number $K=2$. Same as \cite{Wang2019Neural}, we set the node dropout ratio to 0.0, and message dropout ratio to 0.2. We use the cosine similarity with temperature $t>0$ as the similarity function, and set $t_2=0.1$ for Yelp2018 dataset, $t_2=0.07$ for Amazon-Book dataset and Steam dataset. And we set $t_1=0.8$ for all three datasets. We use normalization to stabilize debiased contrastive learning. The calss probability $\tau^{+}$ and loss weight $\beta$ is tuned in $\{0,0.1,0.01,0.001,0.0001,0.00001\}$ and $\{0.1,0.2,0.3,0.4,0.5,0.6,0.7,0.8,0.9,1.0\}$, respectively.  All the models are trained on a single NVIDIA GeForce GTX TITAN X GPU.

\subsection{Overall Performance Comparison (RQ1)}

\begin{table*}[htbp]
	\centering
	\caption{The comparison of overall performance among our methods and competing methods.}
	\begin{tabular}{|c|c|c|c|c|c|c|}
		\hline
		\multirow{3}{*}{{\bf Methods}} & \multicolumn{6}{|c|}{{\bf Dataset}} \\ \cline{2-7}
		& \multicolumn{2}{|c|}{Yelp2018}
		& \multicolumn{2}{|c|}{Amazon-Book}
		& \multicolumn{2}{|c|}{Steam} \\
		\cline{2-7}
		\multicolumn{1}{|c|}{} & \multicolumn{1}{|c|}{recall@20} & \multicolumn{1}{|c|}{ndcg@20} & \multicolumn{1}{|c|}{recall@20} & \multicolumn{1}{|c|}{ndcg@20} & \multicolumn{1}{|c|}{recall@20} & \multicolumn{1}{|c|}{ndcg@20} \\
		\hline
		
		MF & 0.0445 & 0.0361 & 0.0306 & 0.0231 & 0.0632 & 0.0303 \\
		MF++ & $0.0502_{(+12.81\%)}$ & $0.0403_{(+11.63\%)}$ & $0.0489_{\textbf{(+59.80\%)}}$ & $0.0386_{\textbf{(+67.10\%)}}$ & $0.0759_{(+20.09\%)}$ & $0.0390_{(+28.71\%)}$ \\
		\hline
		DMF & 0.0419 & 0.0347 & 0.0268 & 0.0215 & 0.0877 & 0.0441 \\
		DMF++ & $0.0615_{\textbf{(+46.78\%)}}$ & $0.0507_{\textbf{(+46.11\%)}}$ & $0.0408_{(+52.24\%)}$ & $0.0333_{(+54.88\%)}$ & $0.1063_{(+21.21\%)}$ & $0.0547_{(+24.04\%)}$ \\
		\hline
		\hline
		GCN & 0.0457 & 0.0371 & 0.0312 & 0.0242 & 0.0866 & 0.0435 \\
		GCN++ & $0.0649_{(+42.01\%)}$ & $0.0533_{(+43.67\%)}$ & $0.0486_{(+55.77\%)}$ & $0.0381_{(+57.44\%)}$ & $0.1089_{\textbf{(+25.75\%)}}$ & $0.0568_{\textbf{(+30.57\%)}}$ \\
		\hline
		GC-MC & 0.0498 & 0.0411 & 0.0346 & 0.0258 & 0.1023 & 0.0518 \\
		GC-MC++ & $0.0622_{(+24.90\%)}$ & $0.0504_{(+22.63\%)}$ & $0.0457_{(+32.08\%)}$ & $0.0360_{(+39.53\%)}$ & $0.1161_{(+13.49\%)}$ & $0.0605_{(+16.80\%)}$ \\
		\hline
		PinSage & 0.0458 & 0.0369 & 0.0403 & 0.0309 & 0.0849 & 0.0455 \\
		PinSage++ & $0.0479_{(+4.59\%)}$ & $0.0392_{(+6.23\%)}$ & $0.0469_{(+16.38\%)}$ & $0.0369_{(+19.42\%)}$ & $0.1044_{(+22.97\%)}$ & $0.0540_{(+18.68\%)}$ \\
		\hline
		NGCF & 0.0584 & 0.0489 & 0.0365 & 0.0271 & 0.1129 & 0.0597 \\
		NGCF++ & $0.0646_{(+10.62\%)}$ & $0.0532_{(+8.79\%)}$ & $0.0452_{(+23.84\%)}$ & $0.0348_{(+28.41\%)}$ & $0.1149_{(+1.77\%)}$ & $0.0597_{(+1.51\%)}$ \\
		\hline
		\hline
		LR-GCCF & 0.0602 & 0.0493 & 0.0373 & 0.0283 & 0.1035 & 0.0534 \\
		LR-GCCF++ & $\textbf{0.0711}_{(+18.11\%)}$ & $0.0558_{(+13.18\%)}$ & $0.0460_{(+23.32\%)}$ & $0.0340_{(+20.14\%)}$ & $0.1086_{(+4.93\%)}$ & $0.0565_{(+5.81\%)}$ \\
		\hline
		LightGCN & 0.0630 & \underline{0.0519} & 0.0453 & 0.0346 & 0.1159 & 0.0611 \\
		LightGCN++ & $0.0684_{(+8.57\%)}$ & $\textbf{0.0564}_{(+8.67\%)}$ & $0.0506_{(+11.70\%)}$ & $0.0397_{(+14.74\%)}$ & $0.1204_{(+3.88\%)}$ & $0.0628_{(+2.78\%)}$ \\
		\hline
		LightGCN-single & \underline{0.0633} & 0.0506 & \underline{0.0466} & \underline{0.0358} & \underline{0.1246} & \underline{0.0650} \\
		LightGCN-single++ & $0.0690_{(+9.00\%)}$ & $0.0560_{(+10.67\%)}$ & $\textbf{0.0545}_{(+16.95\%)}$ & $\textbf{0.0442}_{(+23.46\%)}$ & $\textbf{0.1382}_{(+10.91\%)}$ & $\textbf{0.0716}_{(+10.15\%)}$ \\
		\hline
	\end{tabular}
	
	\label{tab:comparison}
\end{table*}

Table \ref{tab:comparison} summarizes all competing methods' best results on three benchmark datasets. The percentage in subscript brackets are the relative improvements in our methods to each baseline. Bold scores are the best in each column, and bold percentage in subscript brackets are the best relative improvements, while underlined scores are the second best. We have the following observations:

On the one hand, our framework has the most obvious relative improvement effect on the most basic model such as MF, DMF, and GCN, especially in Yelp2018 dataset, DMF++ is increased by 46.78\% on recall@20 and 46.11\% on ndcg@20. In the Amazon-Book dataset, MF++ outperforms MF 59.80\% on recall@20 and 67.10\% on ndcg@20, respectively. In the Steam dataset, compared with GCN, GCN++ can obtain the relative improvement at 25.75\% and 30.57\% in recall@20 and ndcg@20, respectively. On the other hand, our framework has the best performance on some state-of-the-art models. For example, in the Yelp2018 dataset, LR-GCCF++ reached 0.0711 on recall@20, and LightGCN++ reached 0.0564 on ndcg@20. In Amazon-Book and Steam datasets, LightGCN-single++ consistently outperforms all the other baselines.

For a more fine-grained observation, MF and DMF achieve poor performance on three datasets. This indicates that the inner product is insufficient to capture the complex relations between users and items, further limiting the performance. When we change the BPR loss to GCL+DCL, MF++ and DMF++ have improved to varying degrees on the three datasets, especially in the Amazon-Book dataset, MF++ are respectively 59.82\% and 67.10\% higher than MF on recall@20 and ndcg@20. It is noting that DMF slightly underperforms MF in Yelp2018 and Amazon-Book datasets because DMF not only incorporates explicit ratings and implicit feedback. However, in this paper, we only use implicit feedback. Compared with MF and DMF, GCN and GC-MC obtain a better representation by incorporating the user-item graph's first-order neighbors more fully. By adding graph perturbation and correcting sampling bias, GCN++ and GC-MC++ consistently outperform GCN and GC-MC, respectively. Although the performance of GC-MC on the three datasets is better than GCN, the relative improvement of GC-MC++ is slightly worse than GCN++. It is worth noting that compared to GCN, GCN++ is increased by 42.01\%, 55.77\% and 25.75\% on recall@20 and 43.67\%, 57.44\% and 30.57\% on ndcg@20 in Yelp2018, Amazon-Book, and Steam datasets, respectively. Compared to GCN and GC-MC, PinSage and NGCF further consider the high-order neighbors. More formally, PinSage introduces high-order connectivity in the embedding function, while NGCF can explore the high-order connectivity in an explicit way. Therefore, NGCF generally achieves better performance than PinSage in three datasets. We can see that PinSage++ and NGCF++ consistently outperform PinSage and NGCF, but the relative improvements are slightly worse than GCN++ and GC-MC++.

Compared with the above several methods, LR-GCCF, LightGCN, and LightGCN-single are the current state-of-the-art methods. They not only simplify the complexity of the model but also obtain more competitive performance. Specifically, LR-GCCF removes non-linearities to enhance recommendation performance; LightGCN removes the two most common designs in GCNs - feature transformation and nonlinear activation - degrades the performance of collaborative filtering. LightGCN-single only uses the last layer output for the final prediction. We can find that LR-GCCF++, LightGCN++, and LightGCN-single++ surpass the current best methods in different datasets. Specifically, in Amazon-Book dataset, LightGCN-single++ reaches 0.0545 on recall@20 and 0.0442 on ndcg@20. In Steam dataset, LightGCN-single++ reaches 0.1382 on recall@20 and 0.0716 on ndcg@20.

\subsection{Ablation Study (RQ2)}
\label{sec:ablation}
In this subsection, we focus on LightGCN-single - a state-of-the-art GNN-based recommendation model to analyze. Since there are several components in our framework, we analyze their impacts via an ablation study. Here, we mainly focus on graph perturbation and debiased contrastive learning, as is shown in Figure \ref{fig:ablation}(a) and Figure \ref{fig:ablation}(b). We introduce the variants and analyze their effect, respectively:

\begin{figure}[htbp]
  \centering
  
  \includegraphics[width=\linewidth]{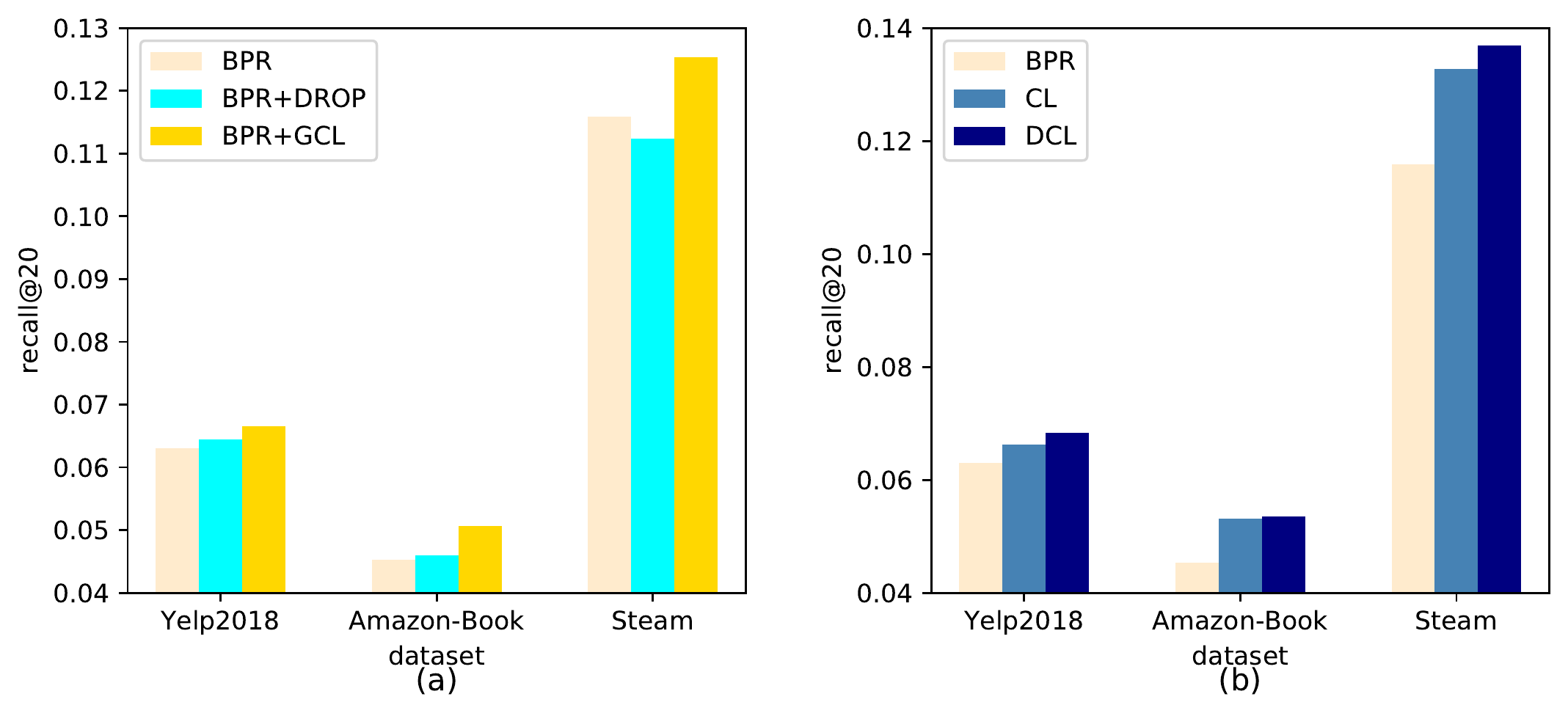}
  \caption{Ablation study (recall@20) on three datasets.}
  \label{fig:ablation}
\end{figure}

\begin{figure*}[htbp]
\centering
\subfigure{
\begin{minipage}[t]{0.32\linewidth}
\centering
\includegraphics[width=\linewidth]{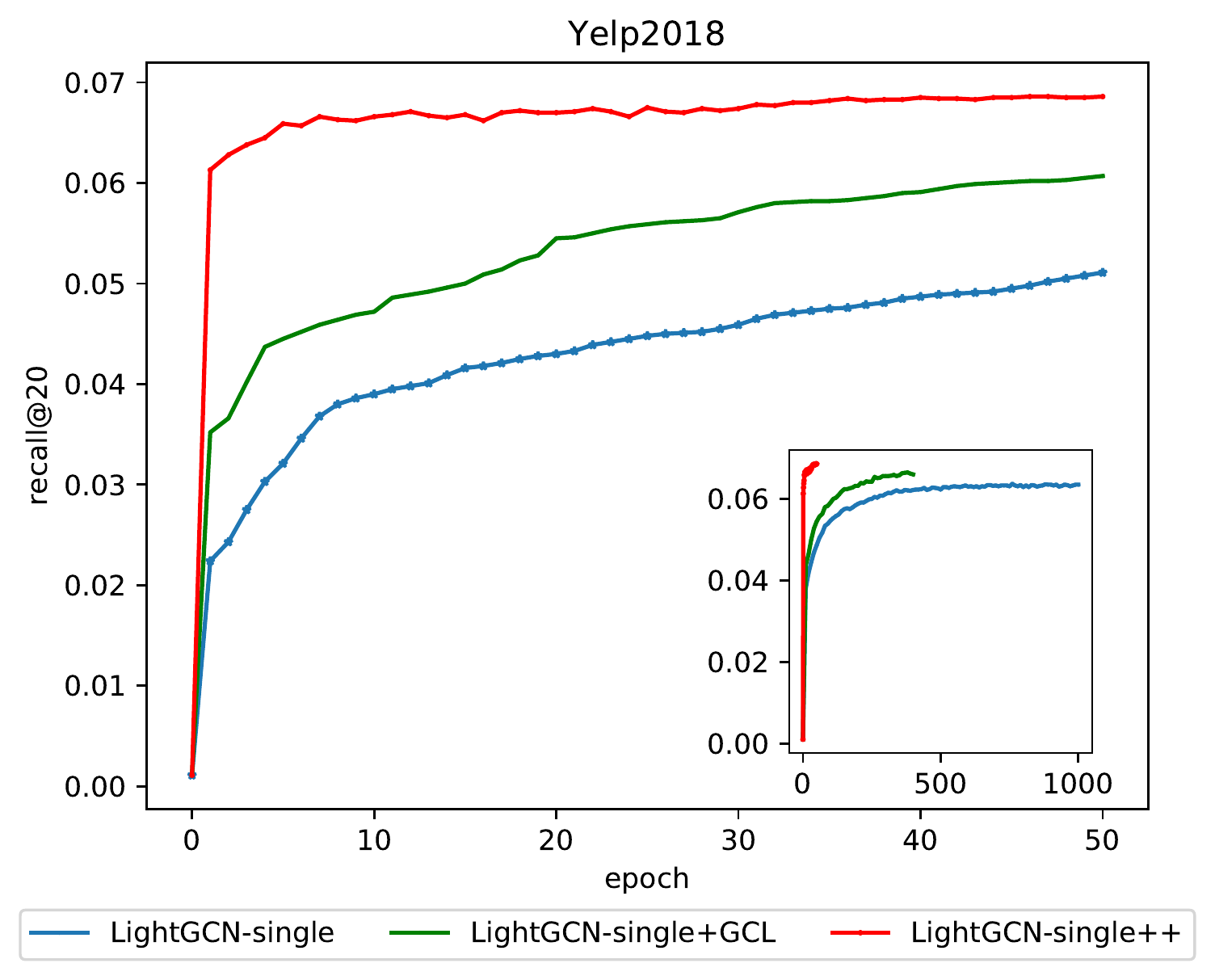}
\end{minipage}%
}%
\subfigure{
\begin{minipage}[t]{0.32\linewidth}
\centering
\includegraphics[width=\linewidth]{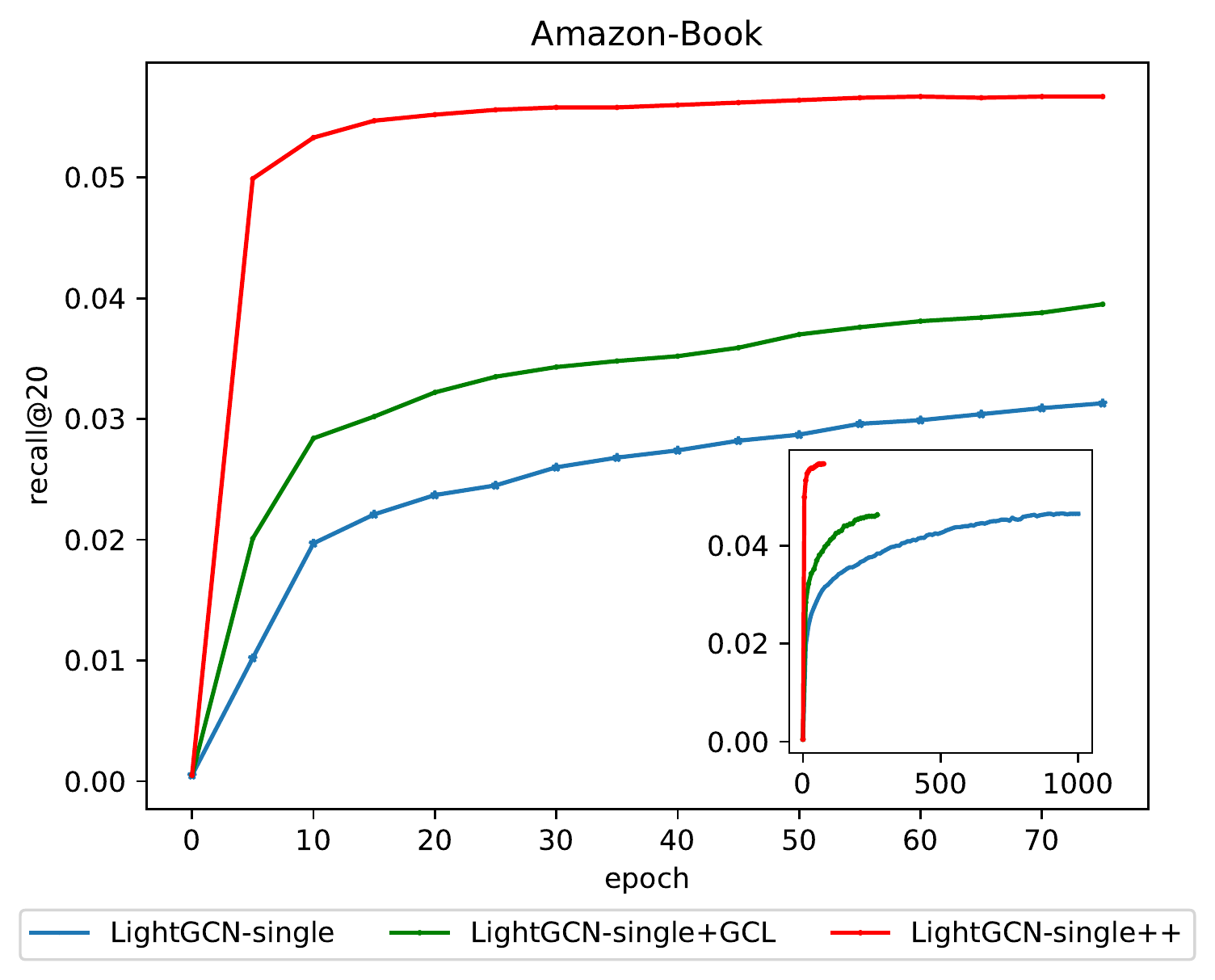}
\end{minipage}%
}%
\subfigure{
\begin{minipage}[t]{0.32\linewidth}
\centering
\includegraphics[width=\linewidth]{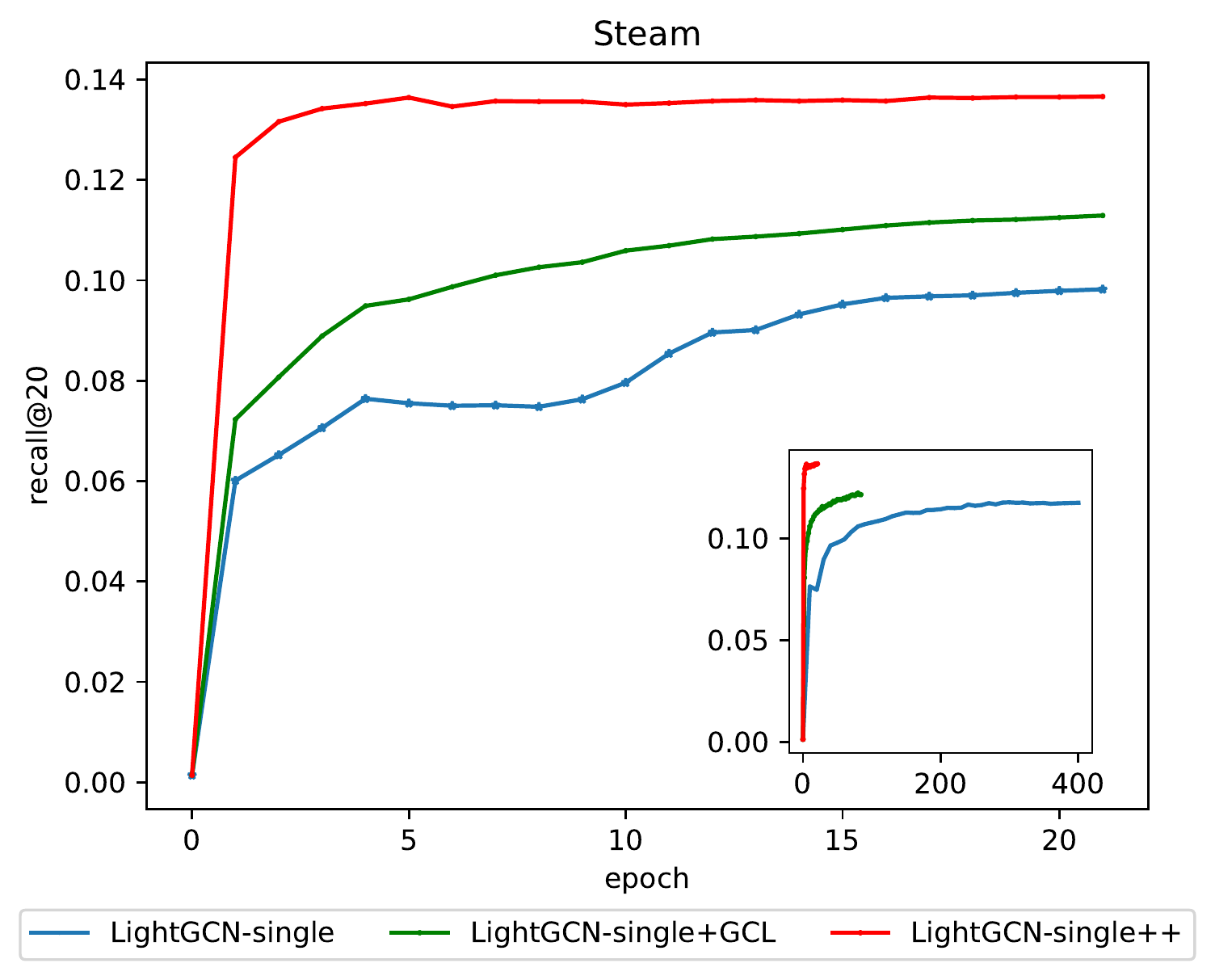}
\end{minipage}
}%
\centering
\caption{Test performance of each epoch of LightGCN-single, LightGCN-single+GCL and LightGCN-single++.}
\label{Fig:effectiveness}
\end{figure*}

\begin{itemize}
\item BPR: We change the loss function to BPR loss, which is a pairwise loss that encourages the prediction of an observed entry to be higher than its unobserved counterparts:
\begin{equation}
    \label{equ:bpr}
    L_{BPR}=-\sum_{u=1}^n\sum_{i\in\mathcal{N}_u}\sum_{j\notin\mathcal{N}_u}\ln\sigma\left(\hat{y}_{ui}-\hat{y}_{uj}\right)+\lambda||\textbf{E}||^2
\end{equation}
where $\lambda$ controls the $L_2$ regularization strength. $\textbf{E}$ is the embedding matrix of all users and items.
\item BPR+DROP: Based on BPR, we add message dropout  \cite{Wang2019Neural} to prevent LightGCN from overfitting.
\item BPR+GCL: Add graph perturbation based on BPR; namely, we apply a stochastic perturbation to the $L$-hop user-item subgraph at each node.
\item CL: It means that the bias correction probability $\tau^+=0$, namely not to correct the sample bias.
\item DCL: Only use the debiased contrastive loss.
\end{itemize}

Figure \ref{fig:ablation} shows the performance each variant on all three datasets. Due to the space limitation, we only report the recall@20, while ndcg@20 has similar trends. We have the following observations:

\textbf{The impact of graph perturbation.} As is shown in Figure \ref{fig:ablation}(a), BPR+GCL generally achieve better performance than BPR. For a more fine-grained analysis, BPR+GCL also outperforms BPR+DROP, because message dropout just randomly drops out the messages, which may be some critical information and deteriorate the performance. For example, in Steam dataset, BPR+DROP slightly underperforms BPR. In comparison, graph perturbation is to learning the embedding of users and items in a self-supervised manner by maximizing the similarity between the representations of two randomly perturbed versions of the link structure of the same node's local subgraph. This reduces randomness to a certain extent and obtains a better representation, which can improve the performance. 


\textbf{The impact of different loss.} As is shown in Figure \ref{fig:ablation}(b), BPR loss achieves poor performance on three datasets since it conducts suboptimal sampling and inefficient use of negative examples. In CL loss, we treat all the other users and items in a minibatch as negative examples that lead to better performance. However, in CL loss, such thousands of negative examples may contain few false negative examples, so we add bias correction probability $\tau^+$ to control sample bias and define it as DCL. We 
find that DCL loss further improves the performance of the recommendation. In summary, adequate negative examples and effective sample bias control are essential to improve the performance of recommendation algorithms.



\subsection{Effectiveness Analysis (RQ3)}

Figure \ref{Fig:effectiveness} shows the test performance w.r.t. recall@20 of each epoch of LightGCN-single, LightGCN-single+GCL and LightGCN-single++. Here, LightGCN-single+GCL means that the loss in LightGCN-single is BPR+GCL. Due to the space limitation, we omit the performance w.r.t. ndcg@20, which has a similar trend with recall@20. Since the convergence speed of the LightGCN-single is relatively slow, we draw a small sub-figure in the lower right corner of each figure. 

We observe that LightGCN-single+GCL exhibits faster convergence than LightGCN-single on all three datasets. More formally, to achieve the best performance, LightGCN-single+GCL requires 300 - 400 epochs in Yelp2018 and Amazon-Book datasets, while LightGCN-single requires even thousands of epochs. In Steam dataset, LightGCN requires about 400 epochs to achieve the best performance, while LightGCN-single+GCL only requires about 70-100 epochs, which has a faster convergence than LightGCN. This shows that graph perturbation accelerates the convergence rate through self-supervision.

We further improve LightGCN-single+GCL, changing the BPR loss to DCL loss, and defining it as LightGCN-single++, which has faster convergence than LightGCN-single+GCL. In Yelp2018 and Amazon-Book datasets, LightGCN-single++ only requires 50 - 70 epochs to achieve the best performance. It is worth noting that it only requires 18 epochs to converge in Steam dataset. It is reasonable since we sample more negative examples for each user, which speeds up the convergence rate. Although we sample thousands of negative samples, this does not increase the memory overhead but directly reuses the examples in current minibatch.

Besides, LightGCN-single+GCL outperforms LightGCN-single, and LightGCN-single++ further outperforms LightGCN-single+GCL. This indicates that graph perturbation and debiased sampling strategy not only improves training efficiency but also improves the recommendation performance. This also provides a new optimization perspective for future recommendation algorithm design.

\subsection{Parameter Sensitivity Analysis (RQ4)}
In this subsection, we examine the robustness of different hyper - parameters, including bias correction probability $\tau^+$ and graph perturbation rate $p$. We analyze one hyper-parameter at a time by fixing the remaining hyper-parameters at their optimal settings.

\subsubsection{Impact of Bias Correction Probability $\tau^+$}
\begin{figure}[htbp]
\centering

\subfigure{
\begin{minipage}[t]{0.32\linewidth}
\centering
\includegraphics[width=\linewidth]{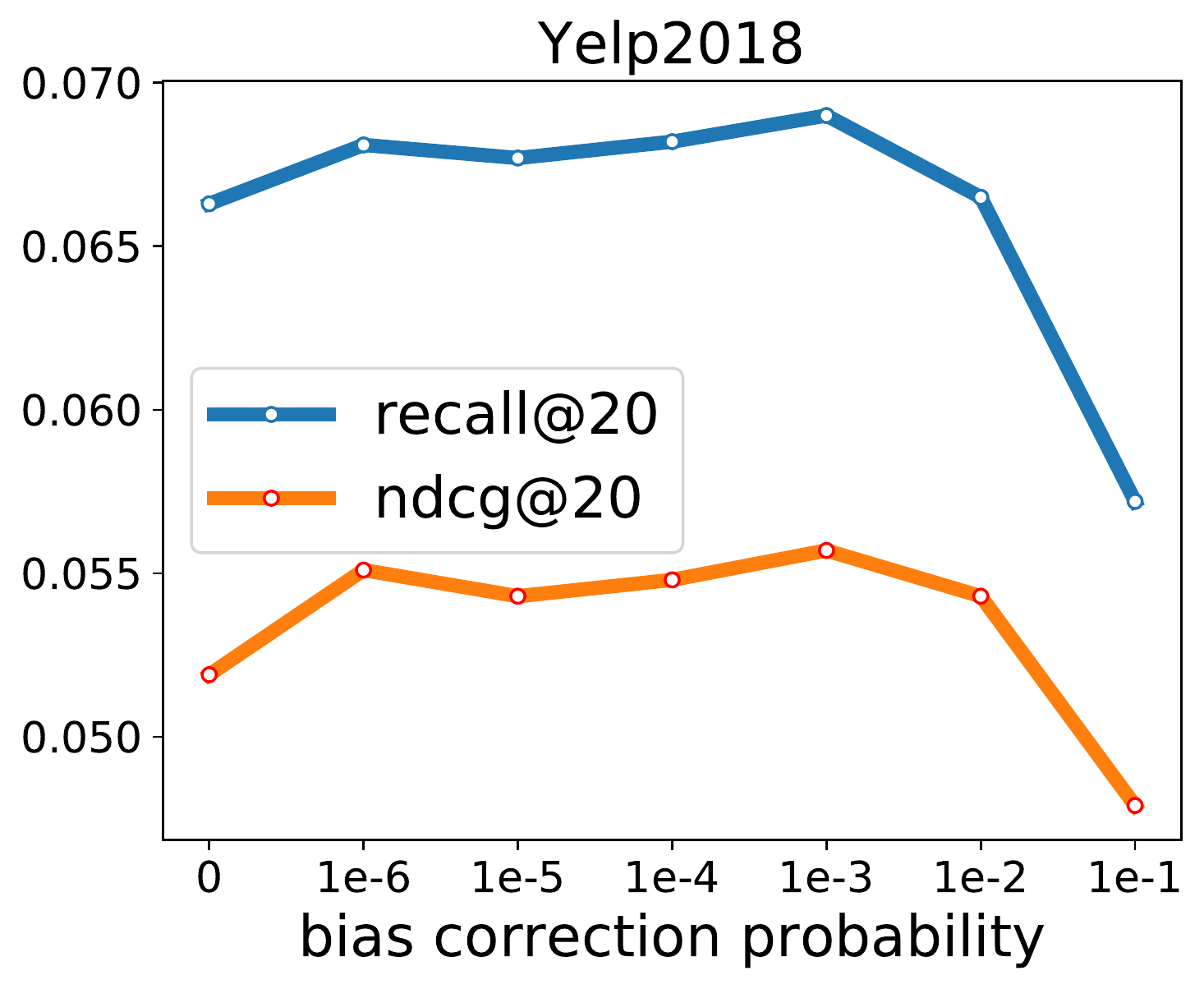}
\end{minipage}%
}%
\subfigure{
\begin{minipage}[t]{0.32\linewidth}
\centering
\includegraphics[width=\linewidth]{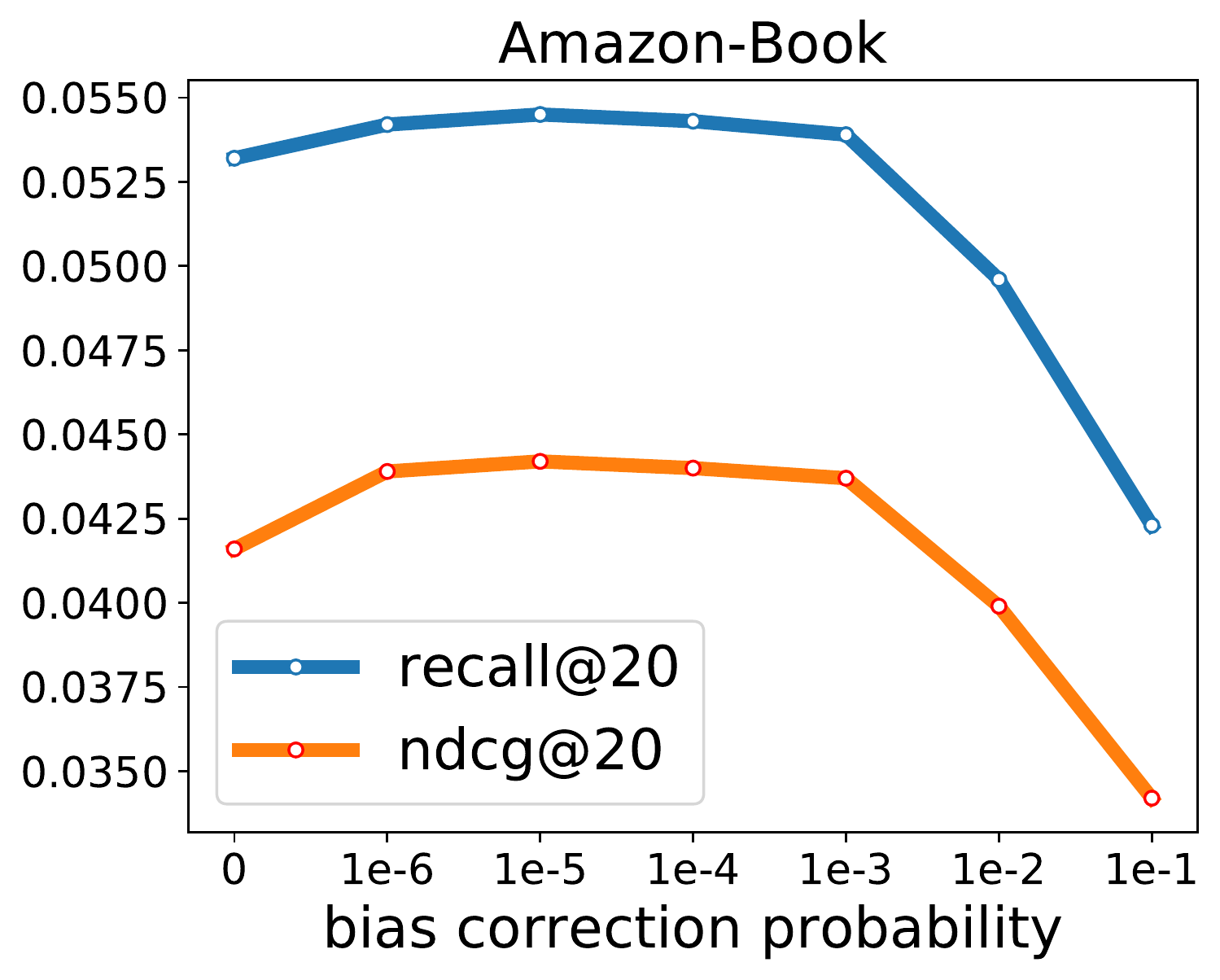}
\end{minipage}%
}%
\subfigure{
\begin{minipage}[t]{0.32\linewidth}
\centering
\includegraphics[width=\linewidth]{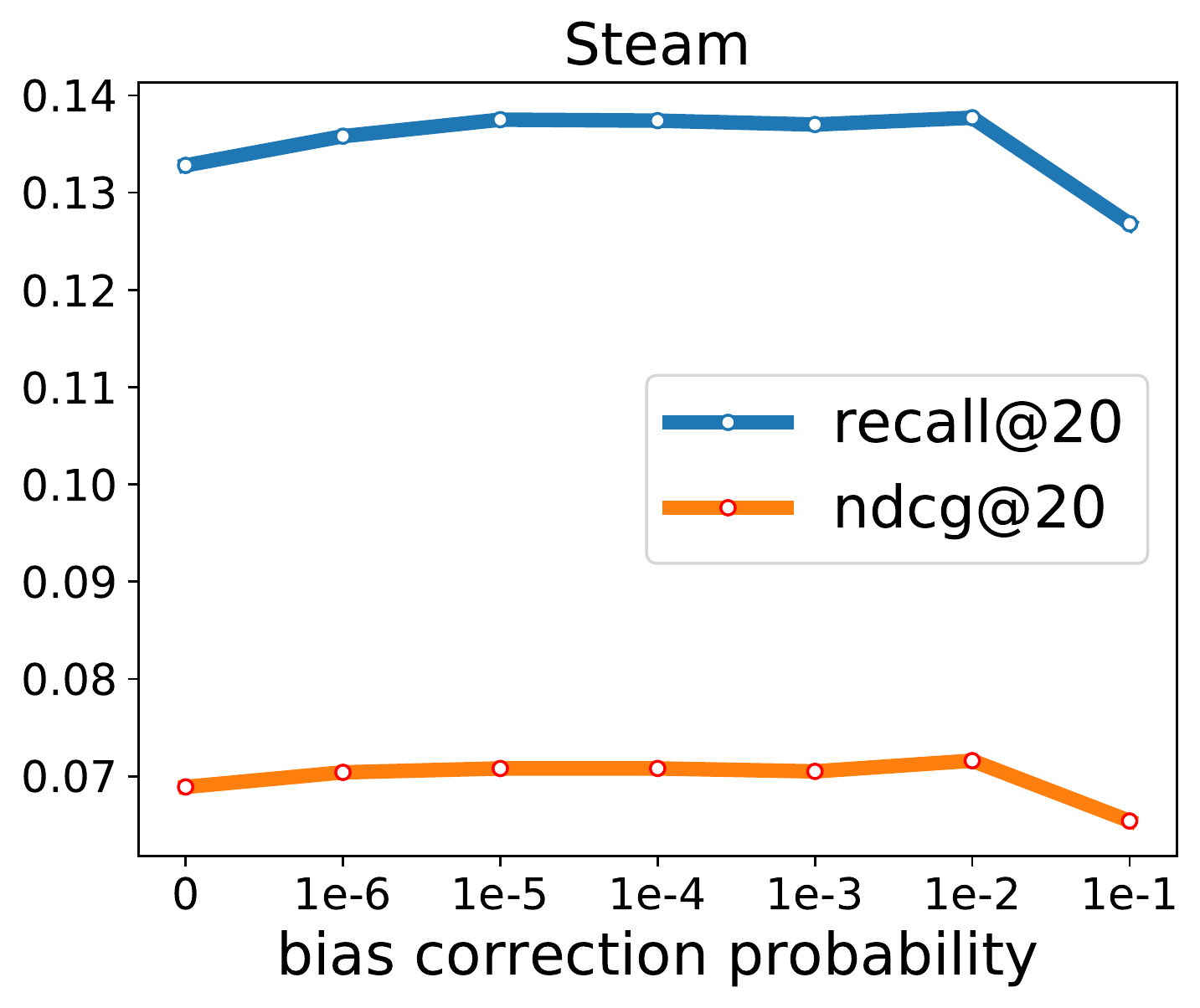}
\end{minipage}
}%
\centering
\caption{Performance of LightGCN-single++ w.r.t. different bias correction probability $\tau^+$ on three datasets.}
\label{Fig:tau_plus}
\end{figure}

We first study how the bias correction probability $\tau^+$ affect the recommendation performance. Figure \ref{Fig:tau_plus} shows recall@20 and ndcg@20 for LightGCN-single++ with the bias correction probability $\tau^+$ varying from 0 to 0.1 while keeping other optimal hyper-parameters unchanged. We make some observations from this figure.

The most obvious observation from these sub-figures is that recall@20 and ndcg@20 have the same trend: they both increase first and then decrease as $\tau^+$ increases. In Yelp2018 and Amazon-Book datasets, recall@20 and ndcg@20 increase steadily, and if we continue to increase $\tau^+$, the performance begins to drop suddenly. While in the Steam dataset, when we vary $\tau^+$ from 0.01 to 0.1, the performance begins to drop. This shows that increasing the value of $\tau^+$ can correct the sampling bias within the tolerable range. When it exceeds this range, that is, when it is considered that these samples are false negative samples to a greater extent, the performance will deteriorate, which is reasonable.

For a more fine-grained analysis, when $\tau^+=0$, it means that we do not correct the sample bias and treat the other $(2B-2)$ users and items in current minibatch as true negative examples. We observe that the performance is relatively weak in these three datasets. It indicates that among these negative examples really exist false negative examples. When we set $\tau^+$ a small value to correct the sample bias, recall@20 and ndcg@20 have improved. We can obtain the best performance when setting $\tau^+$ to $1e^{-3}$, $1e^{-5}$ and $1e^{-2}$ in Yelp2018, Amazon-Book, and Steam datasets, respectively.

\subsubsection{Impact of Graph Perturbation Rate $p$}

\begin{figure}[htbp]
\centering

\subfigure{
\begin{minipage}[t]{0.32\linewidth}
\centering
\includegraphics[width=\linewidth]{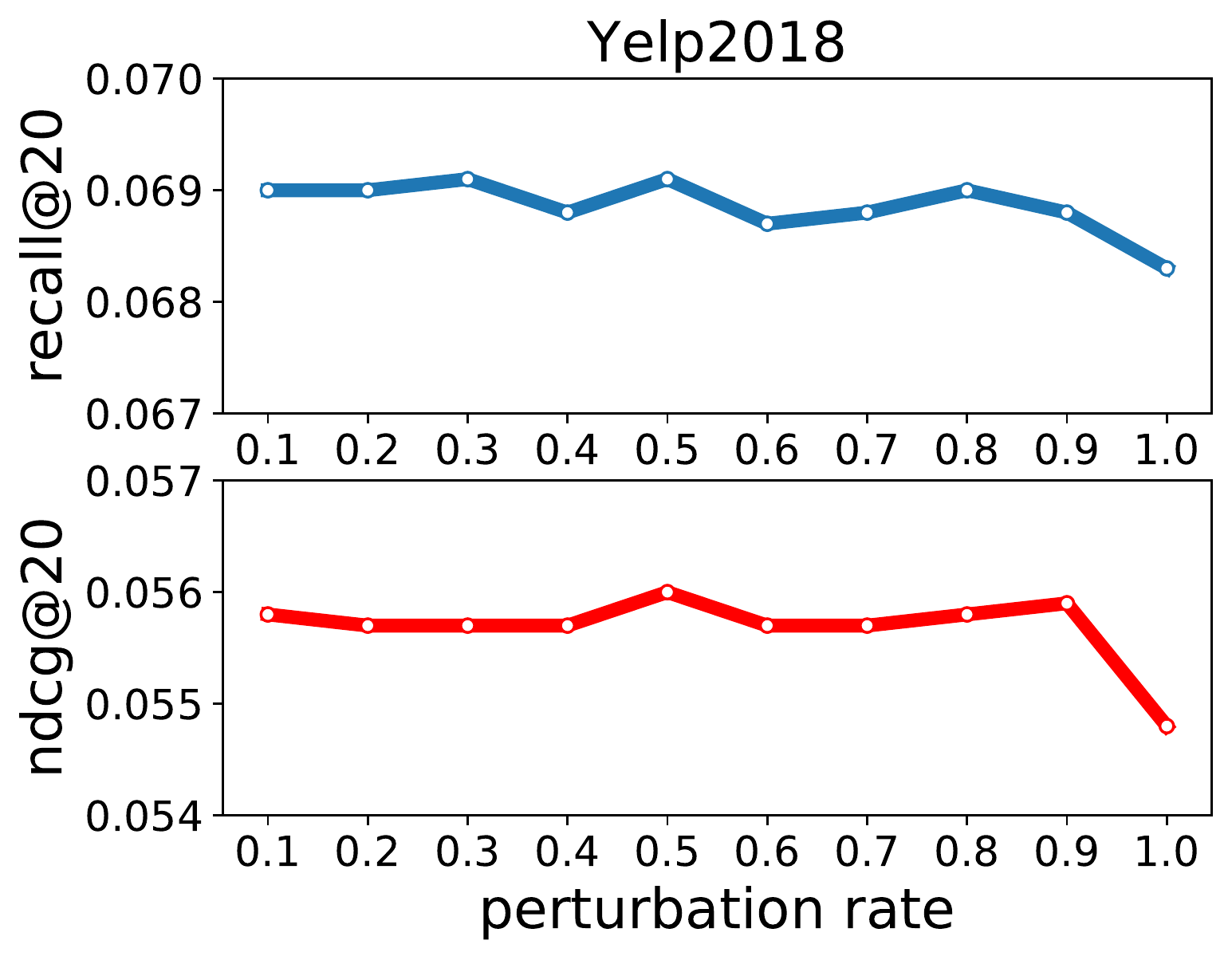}
\end{minipage}%
}%
\subfigure{
\begin{minipage}[t]{0.32\linewidth}
\centering
\includegraphics[width=\linewidth]{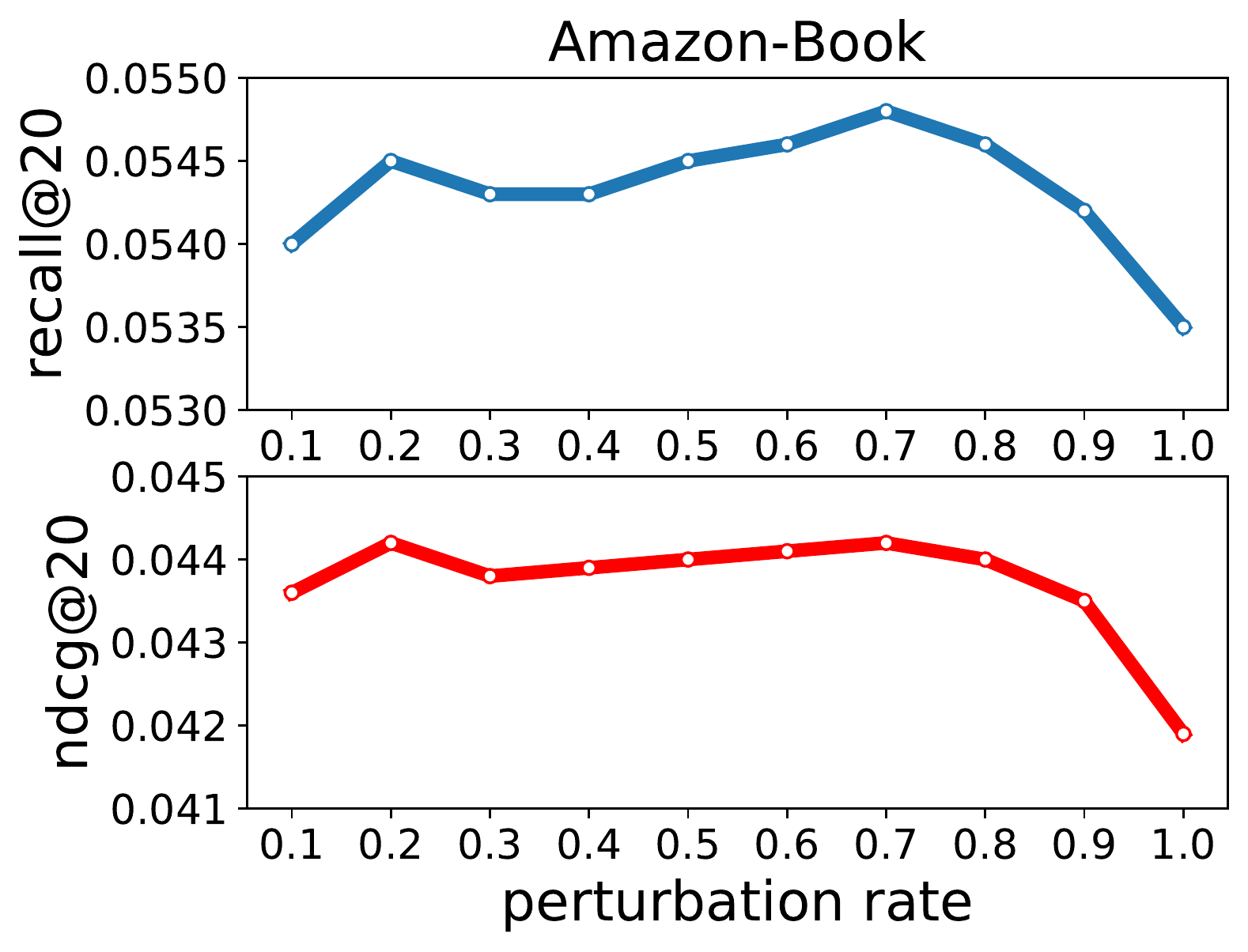}
\end{minipage}%
}%
\subfigure{
\begin{minipage}[t]{0.32\linewidth}
\centering
\includegraphics[width=\linewidth]{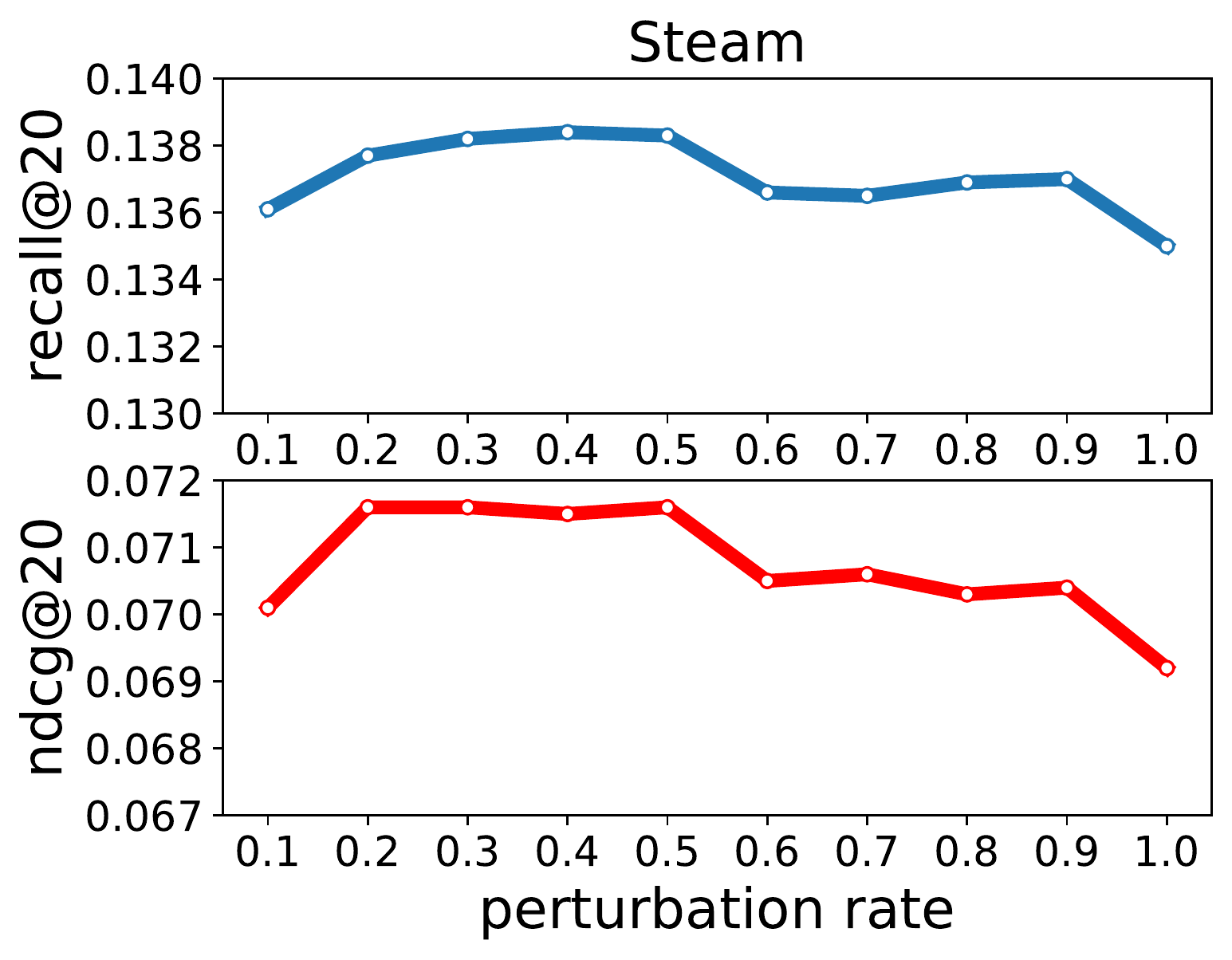}
\end{minipage}
}%
\centering
\caption{Performance of LightGCN-single++ w.r.t. different perturbation $p$ on three datasets.}
\label{Fig:p}
\end{figure}

To study the impact of user-item graph perturbation, we tune the probability $p$ of dropping an edge in $[0.1,1.0]$. The upper part of Figure \ref{Fig:p} shows the performance \emph{w.r.t.} recall@20 for LightGCN-single++, and the lower part shows the performance \emph{w.r.t.} ndcg@20. 

It can be seen that recall@20 and ndcg@20 have the same trend at different perturbation rate. Besides, when the perturbation rate $p$ varies from 0.1 to 0.9, recall@20 and ndcg@20 fluctuate in a small range. This indicates that LightGCN-single++ is robust to different choices of the perturbation parameters $p$. When we set $p$ as 0.3, 0.7, and 0.4 in Yelp2018, Amazon-Book, and Steam datasets, respectively, can result in the best performance.

However, when the perturbation rate $p$ is set to 1.0, recall@20 and ndcg@20 decrease to varying degrees. The reason might be that when $p=1.0$, no graph contrastive loss is added, which fails to obtain a better representation for each user and item. It further shows that graph perturbation has a positive effect on improving the performance of our framework.


\section{Conclusion and  Future Work}
In this work, we present a contrastive learning framework for recommender systems. First, we propose a graph contrastive learning module to reduce the selection bias and obtain more refined embeddings for each user and item. It learns users' and items' representation in a self-supervised manner. Then, to tackle the suboptimal sampling and sample bias, we further develop a debiased contrastive loss, here we treat the other users and items in the current minibatch as negative samples. We provide sufficient negative samples to achieve a satisfying performance. Besides, to alleviate the sampling bias, we use a bias correction probability $\tau^+$ to control the false negative samples. Experimental results on three real-world datasets show that our framework can outperform several general and benchmark recommendation algorithms efficiently.


For future work, we have interests in the following two aspects: on the one hand, we consider applying our model to sequential recommendation tasks. Recent works \cite{DBLP:conf/aaai/WuT0WXT19,DBLP:journals/corr/abs-1910-08887,DBLP:conf/ijcai/XuZLSXZFZ19} apply GNN in sequential recommendation by transforming sequence data into a sequence graph, and achieve great success. However, how to apply the idea of contrastive learning to sequential recommendation tasks and how to construct negative samples to speed up training effectively is still a problem worthy of research. On the other hand, we will explore different ways of graph perturbation. For example, sparse users may retain as much information as possible, while dense users drop their interacted items with a greater probability.

\bibliographystyle{ACM-Reference-Format}

\end{document}